\documentclass[twocolumn,showpacs,pre,english,showkeys,aps,unsortedaddress,superscriptaddress,nofootinbib]{revtex4-1}

\usepackage{amsmath}
\usepackage{amssymb}
\usepackage[colorlinks]{hyperref}
\usepackage{epsfig}
\usepackage{graphicx}
\usepackage[T1]{fontenc}
\usepackage[utf8]{inputenc}
\usepackage[usenames,dvipsnames]{color}
\usepackage[normalem]{ulem}
\usepackage{tikz}
\usetikzlibrary{positioning}

\usepackage{booktabs} 

\usepackage{multirow}
\newcommand{\mr}[2]{\multirow{#1}{*}{#2}}

\begin{document} 

\title{Universal exotic dynamics in critical mesoscopic systems: \\ Simulating the square root of Avogadro's number of spins}

\author{Mauro Bisson$^*$}

\affiliation{NVIDIA Corporation Santa Clara, CA 95051, USA}

\author{Alexandros Vasilopoulos$^*$ }

\affiliation{School of Mathematics, Statistics and Actuarial Science, University of Essex, Colchester CO4 3SQ, United Kingdom}

\author{Massimo Bernaschi}

\affiliation{Istituto per le Applicazioni del Calcolo, CNR - Via dei Taurini 19, 00185 Rome, Italy}

\author{Massimiliano Fatica}

\affiliation{NVIDIA Corporation Santa Clara, CA 95051, USA}

\author{Nikolaos G. Fytas}
\email{nikolaos.fytas@essex.ac.uk}

\affiliation{School of Mathematics, Statistics and Actuarial Science, University of Essex, Colchester CO4 3SQ, United Kingdom}

\author{Isidoro Gonz\'{a}lez-Adalid Pemart\'{i}n}

\affiliation{Istituto per le Applicazioni del Calcolo, CNR - Via dei Taurini 19, 00185 Rome, Italy}

\author{V\'{i}ctor Mart\'{i}n-Mayor}

\affiliation{Departamento de F\'isica
T\'eorica I, Universidad Complutense, 28040 Madrid, Spain}

\def\thefootnote{*}

\date{\today}

\begin{abstract}
We explicitly demonstrate the universality of critical dynamics through unprecedented large-scale GPU-based simulations of two out-of-equilibrium processes, comparing the behavior of spin-$1/2$ Ising and spin-$1$ Blume-Capel models on a square lattice. In the first protocol, a completely disordered system is instantaneously brought into contact with a thermal bath at the critical temperature, allowing it to evolve until the coherence length exceeds $10^{3}$ lattice spacings. Finite-size effects are negligible due to the mesoscopic scale of the lattice sizes studied, with linear dimensions up to $L=2^{22}$ and $2^{19}$ for the Ising and Blume-Capel models, respectively. Our numerical data, and the subsequent analysis, demonstrate a strong dynamic universality between the two models and provide the most precise estimate to date of the dynamic critical exponent for this universality class, $z = 2.1676(1)$. In the second protocol, we corroborate the role of the universal ratio of dynamic and static length scales in achieving an exponential acceleration in the approach to equilibrium just \emph{above} the critical temperature, through a time-dependent variation of the thermal bath temperature. The results presented in this work leverage our CUDA-based numerical code, breaking the world record for the simulation speed of the Ising model.
\end{abstract}

\maketitle

{\it Introduction.}--- Counterintuitive phenomena emerge when a system attempts to reach thermal equilibrium following a temperature change. Examples include remarkable memory and rejuvenation effects in spin glasses~\cite{jonason:98,janus:23,orbach-janus:24}, as well as the Mpemba effect, in which, under certain conditions, the hotter of two identical beakers of water cools faster when placed in contact with a thermal reservoir colder than both~\cite{aristotle-ross:81,jeng:06,mpemba:69,bechhoefer:21} and its variants, see, e.g. Ref.~\cite{gal:20,lapolla:20, teza:25}. Theoretical analysis is often limited to systems in which the contribution of a single time scale is manipulated by varying the temperature of the external bath, leading to surprising effects~\cite{lu:17,teza:23,gonzalez-adalid-pemartin:24}. However, similar phenomena also occur when a continuum of time scales is relevant, such as near a second-order phase transition~\cite{gonzalez-adalid-pemartin:21} or deep inside the spin-glass phase~\cite{jonason:98,janus:19}. It is hoped that universality--the insensitivity of macroscopic behavior to microscopic details~\cite{zinn-justin:05}--may aid the analysis of problems governed by multiple time scales.

The concepts of critical phenomena can, fortunately, be extended to dynamical processes (see Ref.~\cite{hohenberg:77} for a seminal review). However, while universality is well-established for equilibrium properties, its extension to dynamical properties remains less clear and lags behind its theoretical counterpart~\cite{hohenberg:77,folk:06}. A rigorous solution to the critical dynamics of the simplest fruit-fly model in statistical physics—the two-dimensional Ising model~\cite{mccoy:73,newman:99,landau:05}—remains elusive. Moreover, models within the same static universality class do not necessarily belong to the same dynamic universality class~\cite{bonati:25}. In fact, the discussion becomes even more complex for disordered systems~\cite{hasenbusch:07}, where violations of universality have been reported~\cite{daSilva:09,zhong:20}. 

A particularly fundamental and non-trivial case arises in models whose equilibrium critical properties belong to the same universality class as the Ising ferromagnet. A representative example is the spin-$1$ Blume-Capel model~\cite{blume:66,capel:66}, specifically in its second-order transition regime~\cite{zierenberg:17}. As is well known, the onset of criticality is marked by a divergence of both the correlation length $\xi$ and the correlation time $\tau$. While the former divergence yields singularities in static quantities, the latter manifests notably as critical slowing down. To describe dynamical scaling properties, an additional exponent is required in addition to the static exponents. This so-called dynamic exponent $z$ links the divergence of length and time scales, i.e., $\tau \sim \xi^{z}_\mathrm{eq}$~\cite{nightingale:96,nightingale:00,hasenbusch:20,liu:23} ($\xi_\mathrm{eq}$ is the correlation length in equilibrium). In a finite system, $\xi_\mathrm{eq}$ is bounded by the linear system size $L$, so that $\tau \sim L^{z}$ at the incipient critical point. The dynamic critical exponent $z$ has been numerically determined to be $z = 2.1665(12)$ in two dimensions in the seminal work by Nightingale and Blöte~\cite{nightingale:96}, and was later shown to be universal with respect to the underlying lattice structure~\cite{wang:97}. The more recent theoretical studies based on the nonperturbative renormalization-group~\cite{duklut:17} and the $\varepsilon$-expansion~\cite{adzhemyan:22}, suggesting $z \approx 2.15$ and $2.14(2)$, respectively, should also be noted.

However, even the simplest equilibration protocol--where the system is instantaneously quenched from a high temperature to the working temperature $T$, and then allowed to relax for a time $t$--can lead to a range of different outcomes depending on $L$ and $t$. Indeed, the computation of $z$ in Ref.~\cite{nightingale:96} relies on an assumption that is well-founded for finite systems (see, e.g.~\cite{levin:17,sokal:97}), namely that a time scale $\tau$ exists such that \emph{all} observables approach their equilibrium value when $t\approx\tau\propto L^z$. However, in the critical dynamics of a system with $L=\infty$, perfectly reasonable quantities may remain far from their equilibrium values even when $t\gg\tau$ ~\cite{fernandez:18b,fernandez:19}. Moreover,
there are both experimental (see, e.g., Ref.~\cite{orbach-janus:24}) and theoretical~\cite{bray:94} settings in which equilibrium is never fully achieved. In these cases, the size of magnetic domains, $\xi(t)$, grows indefinitely provided $L = \infty$ and $T\leq T_\mathrm{c}$, where $T_\mathrm{c}$ denotes the critical temperature. Precisely at $T=T_\mathrm{c}$, one has $\xi(t) \propto t^{1/z}$, while $\xi(t)\propto\sqrt{t}$ for $T<T_\mathrm{c}$ (if $T$ is very close to $T_\mathrm{c}$ a crossover occurs from critical dynamics at short times to $\xi(t)\propto\sqrt{t}$ at long times). The domain growth may also persist for a long time if $T\gtrsim T_\mathrm{c}$, in which case $\xi(t)$ must grow until it reaches its equilibrium value $\xi_\mathrm{eq}(T)\propto (T/T_{\mathrm{c}}-1)^{-\nu}$ ($\nu = 1$ for the two-dimensional Ising universality class).  It is generally expected that the domain-growth exponent $z$ in this context matches the exponent $z$ observed in the equilibrium, finite-size setting; however, precision tests to confirm this expectation are currently lacking.

In this paper, we demonstrate that universality holds for the critical dynamics of two ferromagnetic models belonging to the same \emph{static} universality class: the spin-$1/2$ Ising ferromagnet and the spin-$1$ Blume-Capel model, within the  $L\gg \xi(t)$ regime. 
The largest simulated system contains $2^{44}\approx 17.6\times 10^{12}$ spins, which is more than the square root of Avogadro's number of spins. Thus, even though we reach $\xi(t)$ values well above $10^{3}$ lattice spacings, our simulations fully represent the thermodynamic limit. In this way, we provide the most accurate determination to date of the exponent $z$, which turns out to be compatible with, but more precise, than the best estimate in the thermal equilibrium regime (i.e. $\xi(t)\gg L$)~\cite{nightingale:96}. We showcase universality not only in the direct-quench settings discussed above, but also in the exponential speed-up achieved through pre-cooling~\cite{gonzalez-adalid-pemartin:21}. Our results were made possible by a CUDA program--which has been made publicly available elsewhere~\cite{bisson:25}--for multi GPU-based simulations, that is an extension of Ref.~\cite{ROMERO2020}, setting a new world record for the simulation speed of the square-lattice Ising model, achieving \( 8.7 \, \text{fs} \) per spin update. The code implements the Metropolis algorithm and exploits three levels of parallelism: multispin coding, checkerboard decomposition, and domain decomposition. From this viewpoint, the current work represents the dawn of a new era in computational statistical physics of lattice spin models. 

{\it Models and physical observables.}--- The Blume-Capel (BC) ferromagnet is described by the Hamiltonian
\begin{equation}
    \mathcal{H}^{\rm (BC)} = -J\sum_{\langle\mathbf{x},\mathbf{y}\rangle} \sigma_{\mathbf{x}}\sigma_{\mathbf{y}}+\Delta \sum_{\mathbf{x}}\sigma_{\mathbf{x}}^2\,,
    \label{eq:Hamiltonian}
\end{equation}
where the spins $\sigma_{\mathbf{x}} = \{-1, 0, +1\}$ are located in the nodes of an $L\times L$ square lattice with periodic boundary conditions, $\langle\mathbf{x},\mathbf{y}\rangle$ indicates summation over nearest neighbors, and $J > 0$ is the ferromagnetic exchange coupling. The parameter $\Delta$ is known as the crystal-field coupling that controls the density of vacancies ($\sigma_{\mathbf{x}} = 0$). For $\Delta\rightarrow -\infty$, vacancies are suppressed and the model becomes equivalent to the simple Ising ($\sigma_{\mathbf{x}} = \pm 1$) ferromagnet. The phase boundary of the Blume-Capel model in the crystal field--temperature plane separates the ferromagnetic from the paramagnetic phase~\cite{zierenberg:17}. At high temperatures and low crystal fields, the ferromagnetic--paramagnetic transition is a continuous phase transition in the Ising universality class, whereas at low temperatures and high crystal fields, the transition is of first order~\cite{blume:66,capel:66}. At zero temperature, the ferromagnetic order prevails and the point $(\Delta_0 = n_{\rm c}J/2, T = 0)$, where $n_{\rm c}$ denotes the coordination number, lies on the phase boundary~\cite{capel:66}. On the other hand, for zero crystal field, the transition temperature is not exactly known. For the present square-lattice model under study, we provide a high-accuracy estimate $T_{\mathrm{c}}^{\text{(BC)}} (\Delta = 0) = 1.6935583(5)$; see Appendix~\ref{sec:BC_Tc}. This result was obtained through a dedicated finite-size scaling analysis, which combines exact results for this universality class~\cite{difrancesco:87,difrancesco:88,salas:00,caselle:02} with extensive Swendsen-Wang simulations~\cite{swendsen:87} on systems with linear sizes up to $L\leq 4096$. For comparison, in the case of the simple Ising model (IM), the exact critical temperature used in the simulations below is $T_{\mathrm{c}}^{\text{(IM)}} = 2/\log(1+\sqrt{2})$.

\begin{figure}
    \includegraphics[width=1.0\linewidth]{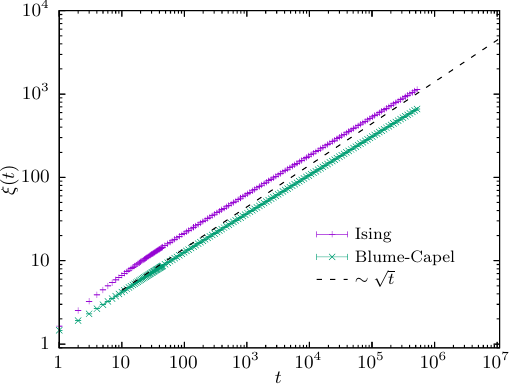}
    \caption{Coherence length $\xi$ as a function of time $t$ for both the Ising ($L = 2^{22}$) and Blume-Capel ($L = 2^{19}$) models at their respective critical points.} 
    \label{fig:xi}
\end{figure}

The main quantities of interest are the correlation function $C(\mathbf{r},t)$ 
\begin{equation}
\label{eq_cor_function}
C(\mathbf{r};t) = \frac{1}{L^2} \sum_{\mathbf{x}-\mathbf{y}=\mathbf{r}}\langle \sigma_{\mathbf{x}}(t)\sigma_{\mathbf{y}}(t)\rangle\,
\end{equation}
and the energy density $E(t)$
\begin{equation}
\label{eq_energy}
E(t) = C(\mathbf{r}=(1,0);t) + C(\mathbf{r}=(0,1);t)\,,
\end{equation}
where $\langle\cdots \rangle$ stands for the average over independent realizations of our thermal protocols, and periodic boundary conditions are understood. We extract the coherence length $\xi(t)$ from space integrals of $C(\mathbf{r};t)$ as explained in Refs.~\cite{janus:08b,fernandez:19b,gonzalez-adalid-pemartin:21}. Interestingly enough, the long- and short-distance
behavior of $C(\mathbf{r};t)$ at $T_\text{c}$ is related through the energy's scaling dimension~\cite{amit:05}, that in our case is $D-1/\nu = 1$. Hence~\cite{fernandez:15, ozeki:07}, at $T_\mathrm{c}$, the energy excess
\begin{equation}\label{eq:E-z}
E(t)-E_\text{eq}\propto [\xi(T_\text{c},t)]^{-(D-1/\nu)}\propto t^{-(1/z)} \,, 
\end{equation}  
where $E_{\rm eq}$ denotes its equilibrium value, provides a short-distance estimate of the dynamic critical exponent $z$. For full details of our simulations and analysis methods, we refer the reader to Appendices~\ref{sec:sim_details}, \ref{sec:cor_function}, and \ref{sec:fittings}.
\begin{figure}
    \includegraphics[width=1.0\linewidth]{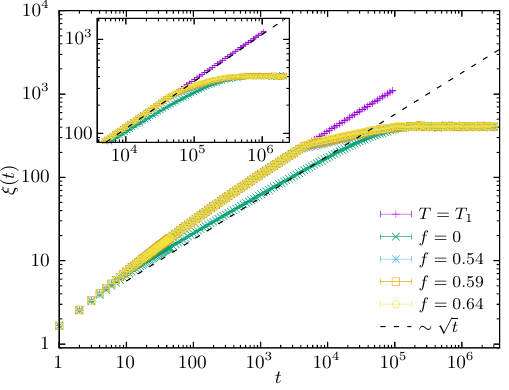}
    \caption{Coherence length $\xi$ for a linear size $L = 2^{16}$ as a function of time $t$ for both the Ising (main panel) and Blume-Capel (inset) models, following the second thermal protocol. $\xi(t)$ appears to be almost $f$-independent in this case.}
    \label{fig:xi_two-step_protocol}
\end{figure}

{\it Thermal protocols.}--- We consider two different thermal protocols. In the direct quench, the system is instantaneously brought from $T = \infty$, (i.e., a fully disordered configuration) to the working temperature $T$ where it is left to relax while the coherence length $\xi(t)$ grows (see Figs.~\ref{fig:xi} and~\ref{fig:xi_two-step_protocol}). The second protocol is a pre-cooling strategy, where a fully disordered configuration is initially placed at a temperature $T_1\approx 0.73 T_{\mathrm{c}}$, allowed to relax for some time, and then heated to a higher temperature $T_2\gtrsim T_{\mathrm{c}}$, where it reaches equilibrium. This protocol is characterized by the fraction $f$ of the equilibrium coherence length $\xi_\text{eq}(T_2)$ that the system reaches while evolving at the lower temperature $T_1$ just before the change to temperature $T_2$, as shown in Fig.~\ref{fig:xi_two-step_protocol}. In Ref.~\cite{gonzalez-adalid-pemartin:21} an exponential dynamic speed-up was observed for the Ising model with $\xi_\text{eq}(T_2)\leq 135$, provided that $f=0.59(7)$. It was speculated that this speed-up is universal. 

\begin{figure}
    \includegraphics[width=1.0\linewidth]{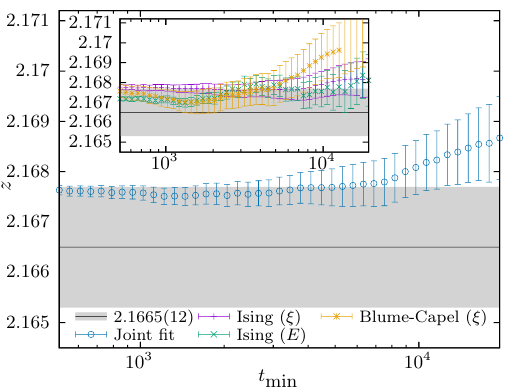}
    \caption{Estimates of the dynamic critical exponent $z$ for the Ising ($L = 2^{22}$) and Blume-Capel ($L = 2^{19}$) models obtained via non-local [$\xi$, Eq.~\eqref{eq:z}] and local [$E$, Eq.~\eqref{eq:z_for_E}] observables, for a varying fitting window ($t_{\text{min}}$: shortest time included in the fit). The main panel displays results from a joint fit for both models (considering data from $\xi$ and $E$ for the Ising model, and only $\xi$ for the Blume-Capel model), as indicated in the legend. The inset shows three sets of data points corresponding to individual fits for each model. We have  acceptable values of the fits' figure of merit $\chi^2/\text{dof}$, where ${\rm dof}$ stands for the number of degrees of freedom, for all the $t_{\text{min}}$ shown. In both panels, the solid line with the gray-scale background
    corresponds to $z = 2.1665(12)$~\cite{nightingale:96}.} 
    \label{fig:z}
\end{figure}

{\it Results.}--- We determine the dynamic critical exponent $z$ from our direct-quench simulations for both models at their respective critical points, see Fig.~\ref{fig:z}. To validate universality, we perform fits of $\xi(t)$ of the form 
\begin{equation}
\label{eq:z}
    \log{(t)} = z \log{[\xi(t)]} + b + b'\xi(t)^{-\omega},
\end{equation}
across the data from both models. Here $b$ and $b'$ are non-universal fitting constants and $\omega = 1.75$ is the corrections-to-scaling exponent for the Ising universality class (see, for example, the discussion in the supplementary material of Refs.~\cite{shao:16,nienhuis:82}).
Furthermore, we compute $z$ also from the energy $E(t)$, as outlined in Eq.~\eqref{eq:E-z}, exclusively for the Ising model, since only in this case is the equilibrium energy value exactly known [$E_\text{eq}(T_{\text{c}})=-\sqrt{2}$]. We consider a fit of the form
\begin{equation}
\label{eq:z_for_E}
    E(t) - E_{\text{eq}}= bt^{-(1/z)} + b't^{-(2/z)} + b''t^{-[(\omega+1)/z]},
\end{equation}
where again $b$, $b'$, and $b''$ are non-universal fitting constants. Indeed, for a system of size $L$ in \emph{equilibrium}, one has $\langle E_L\rangle+\sqrt{2}=c_1/L +{\cal O}(L^{-2})$, where $c_1$ is a known constant~\cite{ferdinand:69}. Dynamic scaling asserts that we can apply the equilibrium result by replacing $L$ with $\xi(t)$ (noting that amplitudes, such as $c_1$, may differ from their equilibrium values). To ensure our determination of $z$ from the energy is independent of long-distance observables, we have further replaced $\xi(t)$ with $t^{1/z}$.  Additionally, we have included a correction term, $t^{-[(\omega+1)/z]}$, to account for scaling corrections in $\xi(t)$, as described in Eq.~\eqref{eq:z} that we have inverted as $\xi=B_0 \,t^{1/z} (1+B_1/t^{\omega/z} +\cdots)$ where $B_0$ and $B_1$ are scaling amplitudes.

Our fitting results are documented in Fig.~\ref{fig:z}. In particular, the main panel presents the dynamic exponent $z$, which we obtained from a joint fit across three data sets, as indicated in the legend. The estimates for $z$ fall within the range $2.16763(9) - 2.1678(5)$ and we report the final result as
\begin{equation}
\label{eq:z_result}
z = 2.1676(1),
\end{equation}
which corresponds to the logarithmic midpoint of the stable region shown in the main plot. This value is in excellent agreement with the result of $z = 2.1665(12)$ reported by Nightingale and Bl{\"o}te~\cite{nightingale:96}. Notably, the error in our calculation is approximately one-twelfth of the error reported in Ref.~\cite{nightingale:96}. In the inset of Fig.~\ref{fig:z}, we show the individual $z$ estimates obtained from separate fits to each of the three data sets, which further support the validity of both the joint fit and the result in Eq.~(\ref{eq:z_result}). For more details about the fits and our statistical accuracy criteria, we refer to Appendix~\ref{sec:fittings}. Note also that the absence of finite-size effects for $\xi(t)$ and $E(t)$ (within our time window and statistical errors) was assessed by comparing results for the Ising model on systems with $L=2^{20}$ and $2^{22}$ linear dimensions (see Appendix~\ref{sec:thermo_limit} for details).

\begin{figure}
    \includegraphics[width=1.0\linewidth]{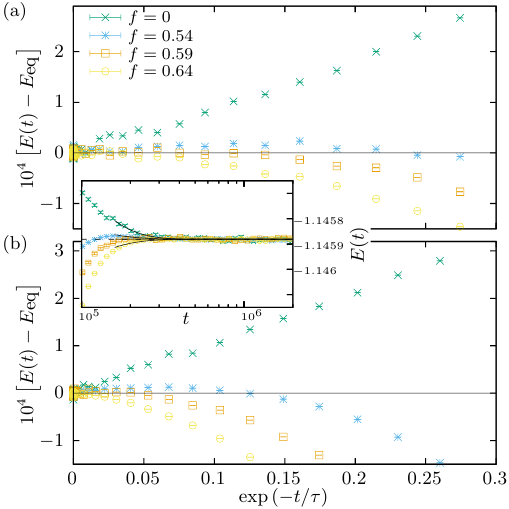}
    \caption{Energy's convergence to its equilibrium value ($E_{\rm eq}$) within the two-step thermal protocol scheme for the (a) Ising and (b) Blume-Capel models. We recall that, for the case of the Ising ferromagnet, $E_{\rm eq}$ is known exactly from the Onsager solution. The inset highlights $E(t)$ in the large-time window for the Blume-Capel case. Results for $L = 2^{16}$ are shown.}
    \label{fig:energy_two-step_protocol}
\end{figure}

Next, we focus on the approach to equilibrium in the paramagnetic phase for $T_2\gtrsim T_{\mathrm{c}}$. Specifically, we choose $T_2$, such that the equilibrium values of $\xi$ are $\xi_\text{eq}(T_2)=407(1)$ for the Ising model and $405(2)$ for the Blume-Capel model. We then examine the approach to equilibrium of the energy under different thermal protocols, as shown in Fig.~\ref{fig:energy_two-step_protocol}. The different protocols maintain $T_1\approx 0.73 T_{\mathrm{c}}$ until $\xi(t,T1)=f \xi_\text{eq}(T_2)$, then the system is instantaneously placed in contact with the thermal bath at $T_2$. The direct quench, where $f=0$, is shown in Fig.~\ref{fig:energy_two-step_protocol}. In the case of direct quench, 
$E(t)$ approaches equilibrium monotonically from above, whereas the three pre-cooling strategies begin well below the asymptotic equilibrium value. However, $E(t;f=0.54)$ overshoots the equilibrium value of the energy and approaches equilibrium from \emph{above}, while $E(t;f=0.59)$ and $E(t;f=0.64)$ approach equilibrium from \emph{below}. An exponential law turns out to aptly fit the late approach to equilibrium
\begin{equation}
    \label{eq:two_step_protocol}
    E(t;f)=E_{\rm eq} + b(f)\exp{(-t/\tau)},
\end{equation}
where the fitting parameters for the Ising model are the amplitude $b(f)$ and the time scale $\tau$, with $E_{\rm eq}$ known from Onsager's solution. We perform a joint fit to the data for all values of $f$, because $\tau$ is $f$-independent in Eq.~\eqref{eq:two_step_protocol}. For the Blume-Capel model, which lacks an exact solution, we have to include $E_{\rm eq}$ as an $f$-independent parameter in the joint fit. In this way, we find for both models $b^{\text{IM}}(f=0.59)=0.00000(2)$, $b^{\text{BC}}(f=0.59)=0.00000(6)$. Furthermore, the amplitudes are positive at $f=0.54$ [$b^{\text{IM}}(f=0.54)=0.00015(3)$, $b^{\text{BC}}(f=0.54)=0.00028(6)$] and negative at $f=0.64$
[$b^{\text{IM}}(f=0.64)=-0.00024(4)$, $b^{\text{BC}}(f=0.64)=-0.0005(1)$]. Therefore, we propose the conservative estimate $f=0.59(5)$ as the universal value at which the exponential speed-up in the approach to equilibrium occurs in the scaling limit for both the Ising and Blume-Capel models.

{\it Discussion.}--- Although there are strong theoretical reasons to expect universality in dynamical critical phenomena~\cite{hohenberg:77, zinn-justin:05}, high-accuracy computations confirming this are still scarce. Moreover, dynamics opens up a broader range of questions. For instance, it is not at all obvious that the limits of large system size and long times are interchangeable. The computation in Ref.~\cite{nightingale:96} relies on the assumption that a time scale 
$\tau$ exists such that \emph{all} physical quantities approach their large-time limits with corrections of the form $\sim\mathrm{e}^{-t/\tau}$. This assumption holds if the limit of large times is taken \emph{before} the thermodynamic limit~\cite{sokal:97,levin:17}, but it is simply not true when the order of limits is reversed, which is the relevant case for field-theoretical computations~\cite{duklut:17, adzhemyan:22}. Furthermore, the temperature of the thermal bath can be varied in time in clever ways to produce counterintuitive effects (see, e.g., Refs.~\cite{aristotle-ross:81, jeng:06, mpemba:69, bechhoefer:21, gal:20, lapolla:20, lu:17, teza:23, gonzalez-adalid-pemartin:24, gonzalez-adalid-pemartin:21, janus:19, teza:25}). Demonstrations of universality in this broader context of time-varying temperatures are rare (if at all existing), though one might argue that memory and rejuvenation effects in spin-glasses provide another example of universality~\cite{janus:23}.

In this work, we have made progress on both fronts, thanks to massive-scale simulations on GPUs. Building on ideas presented in Refs.~\cite{ROMERO2020,bernaschi:24}, we developed a CUDA code that enabled us to simulate square lattices with side lengths $L=2^{22}$ for the Ising model and $L=2^{19}$ for the Blume-Capel model, reaching coherence lengths well over $10^{3}$ lattice spacings. Thus, while we are taking the limit of large times, the limit of large sizes is certainly taken \emph{before}. Through this, we have explicitly demonstrated universality by showing that both models are governed by the same dynamic exponent $z$. Along the way, we have obtained the best estimate to date for the dynamic critical exponent, setting a new standard for comparison with field-theoretical computations. Additionally, we considered the pre-cooling strategy from Ref.~\cite{gonzalez-adalid-pemartin:21}, which achieves an exponential speed-up in the approach to equilibrium at temperatures $T\gtrsim T_\mathrm{c}$ by temporarily entering the ferromagnetic phase. Both the Ising and Blume-Capel models exhibit exponential speed-up when the dimensionless ratio of coherence lengths reaches a specific value, a hallmark of a \emph{dynamic universality class}. We anticipate that the numerical approach developed in this study may be adapted to help resolve ongoing debates about the universality classes of other nonequilibrium systems, including those encountered in active matter~\cite{chen:25}.

\footnotetext{These authors contributed equally to this work.}

\begin{acknowledgments}
Part of the numerical calculations reported in this
paper were performed at the High-Performance Computing
cluster CERES of the University of Essex. This work was partially supported by MCIN/AEI/10.13039/501100011033 and by
``ERDF A way of making Europe'' through Grant No. PID2022-136374NB-C21. The work of AV and NGF was supported by the Engineering and Physical Sciences Research Council (grant EP/X026116/1 is acknowledged).
\end{acknowledgments}

\appendix

\section{Simulation details}
\label{sec:sim_details}

The key details of our simulations are presented in Tables~\ref{tab:one-T} and~\ref{tab:two-Ts}. As discussed in the main text and in Sec.~\ref{sec:BC_Tc}, all simulations related to the Blume-Capel model were conducted at $\Delta=0$.
\begin{table*}[ht!]
\setlength{\tabcolsep}{1.0em}
\begin{tabular}{ l c c c r r }
\toprule \toprule
               &                $L$ &                     $T_{\rm c}$      &      $t_{\rm max}$ &   \#runs & GPU hours \\ \cmidrule{1-6}
 \mr{2}{Ising} &  $2^{20\dagger}$ & \mr{2}{2.26918531421302196814} & \mr{2}{524288} &       80 &     10269 \\ 
               & $2^{22\ddagger}$ &                                &                &       40 &     61687 \\ \cmidrule(lr){1-6} 
  Blume-Capel  &  $2^{19\dagger}$ &         1.69355839821214427017 &         524288 &      160 &      8060 \\
\bottomrule \bottomrule
\multicolumn{6}{l}{\scriptsize$^{\dagger}$Runs performed on either 8xH100 or 4xGH200 NVIDIA GPUs.}\\
\multicolumn{6}{l}{\scriptsize$^{\ddagger}$Runs performed on 64xGB200 GPUs of an NVL72 Multi-Node NVLink system.}
\end{tabular}
\caption{{\bf Simulations at a single temperature:} All simulations were conducted using our best estimate for $T_{\rm c}$, which is precisely known for the Ising model from the Onsager solution. For the Blume-Capel model, this estimate was obtained as described in Sec.~\ref{sec:BC_Tc}. In each case, we report the lattice size $L$, our estimate of $T_{\rm c}$, and the maximum number of Metropolis full-lattice sweeps $t_{\rm max}$. Additionally, we provide the total number of GPU hours, calculated as the sum of the wallclock time multiplied by the number of GPUs used for each run in a set.}\label{tab:one-T}
\end{table*}

\begin{table*}[ht!]
\setlength{\tabcolsep}{1.0em}
\begin{tabular}{ l c c c c c r r }
\toprule \toprule
              &                $L$ &  $T_1$ & $t_{\rm switch}$ &               $T_2$       &       $t_{\rm max}$ &      \#runs &   GPU hours \\ \cmidrule{1-8}
\mr{3}{Ising} & \mr{3}{$2^{16}$} & \mr{3}{1.66} &         5312 & \mr{3}{2.27414861561203} & \mr{3}{3000000} & \mr{3}{200} & \mr{3}{607} \\
              &                  &              &         4467 &                          &                 &             &             \\  
              &                  &              &         3746 &                          &                 &             &             \\ \cmidrule(lr){1-8}     
\mr{3}{Blume-Capel}    & \mr{3}{$2^{16}$} & \mr{3}{1.24} &        48599 & \mr{3}{1.69661396797728} & \mr{3}{2097152} &         300 &         978 \\  
              &                  &              &        41270 &                          &                 &         200 &         652 \\  
              &                  &              &        34540 &                          &                 &         400 &        1302 \\ \cmidrule(lr){1-8}
\mr{2}{Blume-Capel}    & \mr{2}{$2^{16}$} & \mr{2}{1.24} &        48599 & \mr{2}{1.69661396797728} &  \mr{2}{524288} &        1600 &        1317 \\
              &                  &              &        34540 &                          &                 &        4000 &        3294 \\
\bottomrule \bottomrule
\end{tabular}
\caption{{\bf Simulations at two temperatures:} A completely disordered system (i.e., initially at $T=\infty$) began to evolve at a temperature $T_1<T_{\rm c}$, until time $t_{\rm switch}$, when the temperature was changed to $T_{\rm 2}$, and the system continued to evolve until the final time step, $t_{\rm max}$. The remaining notational conventions follow those in Table~\ref{tab:one-T}. For the Blume-Capel model, we first conducted an exploratory set of runs on a large timescale ($\sim 2\times 10^{6}$). Based on the resulting statistics, we then performed additional runs, focusing on the cases that required more data.
All runs were performed on 4xGH200 GPUs.}\label{tab:two-Ts}
\end{table*}

\section{Critical temperature of the Blume-Capel model}
\label{sec:BC_Tc}

The phase boundary of the model in the crystal field-temperature ($\Delta$, $T$) plane separates the ferromagnetic and paramagnetic phases. The critical line $T_\mathrm{c}(\Delta)$ for $0 \leq \Delta < \Delta_{\rm t}$, where $\Delta_{\rm t} = 1.9660(1)$~\cite{kwak:15} is the $\Delta$ coordinate of the tricritical point, has been extensively studied in previous numerical works for the square lattice~\cite{silva:06,malakis:10,kwak:15,zierenberg:17}. The exact value of the Ising limit for the critical line ($\Delta \to -\infty$), given by $T_{\mathrm{c}}(\Delta \to -\infty) = 2/\log(1 + \sqrt{2})$, is well known from Onsager's solution. A consistent trend in the literature indicates that $T_{\mathrm{c}}(\Delta)$ decreases as $\Delta$ increases. This observation motivated our choice of working specifically at $\Delta = 0$. By doing so, we aimed to be as far as possible from the Ising $\Delta \rightarrow \infty$ limit while maintaining a high temperature range. We chose the high-temperature regime because the random number generator used in our CUDA program relies on 32-bit integers~\cite{bisson:25}. In this context, $\Delta = 0$ emerged as a reasonable trade-off, and we set out to improve the accuracy of the known value $T_c(\Delta = 0) = 1.693(3)$~\cite{malakis:10}. To achieve this, we applied finite-size scaling under thermal equilibrium conditions~\cite{barber:83,cardy:12,amit:05}.

We generated equilibrium configurations of the Blume-Capel model at $\Delta = 0$ on square lattices with linear sizes $L = 2^n$ where $5 \leq n \leq 12$, using periodic boundary conditions. To achieve equilibration, we combined Swendsen-Wang cluster updates~\cite{swendsen:87} (restricted to spins with $\sigma_{\mathbf{x}} = \pm 1$) with Metropolis updates that allow for variations in the value of $\sigma_{\mathbf{x}}^2$. Our elementary Monte Carlo step (EMCS) in this equilibrium simulation consisted of $10$ full-lattice Metropolis sweeps followed by one iteration of the Swendsen-Wang algorithm. Even for the largest lattice, $L = 4096$, we found that $40$ EMCS were sufficient to bring the system to thermal equilibrium. This was verified by comparing simulations starting from a fully random configuration with those starting from the ground state of the Hamiltonian (i.e., $\sigma_{\mathbf{x}} = 1$ for all lattice sites ${\mathbf{x}}$). Histogram reweighting was used to extrapolate the results obtained at the simulation's inverse temperature $\beta$ to neighboring values~\cite{falcioni:82,ferrenberg:88}. We simulated $100$ independent runs for each lattice size to increase statistical accuracy at the High-Performance Computing cluster CERES of the University of Essex. Each independent run consisted of $4\times 10^5$ EMCS (measurements were taken every 40 EMCS). Although it is certainly an overkill, the first 10\% of each independent run was discarded to ensure equilibration.

From these equilibrium simulations, we computed the Binder cumulant $U_4$ and the second-moment correlation length $\xi_2$. Both quantities were obtained via the Fourier transform of the spin-field $\mathcal{M}$ and the corresponding susceptibilities $\chi$:
\begin{align}
    {\cal M}(\mathbf{k})=\sum_{\mathbf{x}}\, {\mathrm{e}}^{\mathrm{i}\mathbf{x}\cdot\mathbf{k}}
    \sigma_{\mathbf{x}}\,,\quad \chi(\mathbf{k})=\frac{1}{L^2}\langle |{\cal M}(\mathbf{k})|^2\rangle\,,\quad \nonumber \\ \mathbf{k}=\frac{2\pi}{L}(n_x,n_y)\,,\ n_x,n_y=0,1,\ldots,L-1\,, 
\end{align}
and
\begin{equation}
    U_4=\frac{\langle {\mathcal M}^4(\mathbf{0})\rangle}{\langle {\mathcal M}^2(\mathbf{0})\rangle^2}\,,\quad \xi_{2}=\frac{1}{2\sin(\pi/L)}\sqrt{\frac{\chi(\mathbf{0})}{\chi_{k_{\rm min}}}-1}\,,
\end{equation}
where we have defined $\chi_{k_{\rm min}}=\chi(2\pi/L,0)=\chi(0,2\pi/L)$. 
At the critical temperature (for convenience, we use the inverse temperature $\beta = 1/T$), $U_4$, and $\xi_2/L$ reach universal large-$L$ limits which are known to very high accuracy~\cite{salas:00}:
\begin{align*}
   U_4^*\equiv\lim_{L\to\infty} U_4(L,\beta=\beta_\mathrm{c})= 1.1679229(47)\,,\quad \\
   \left[\frac{\xi_2}{L}\right]^*\equiv \lim_{L\to\infty}\frac{\xi_2(L,\beta=\beta_{\text{c}})}{L}=0.9050488292(4)\,.
\end{align*}
We note here that Salas and Sokal started from the results of Refs.~\cite{difrancesco:87,difrancesco:88} and managed to express both limits in terms of low-dimensional integrals that were evaluated numerically. The term \emph{universal} here means that any model in the universality class of the two-dimensional Ising model, and specifically the Blume-Capel model at $\Delta = 0$, should exhibit these values for the corresponding quantities on a square lattice. Therefore, we identified the inverse temperatures $\beta_{L,U}$ and $\beta_{L,\xi_2/L}$ that satisfy:
\begin{equation}
    U(L,\beta_{L,U})=U^{\ast}\, \text{ and } \frac{\xi_2(L,\beta_{L,\xi_2/L})}{L}=\left[\frac{\xi_2}{L}\right]^{\ast}.
\end{equation}
Histogram reweighting extrapolations were essential to achieve this goal. As noted in Ref.~\cite{amit:05}, dimensionless quantities like $g$ (both $U_4$ and $\xi_{2}/L$ are examples) are expected to scale according to $g(L,\beta)=f_0[L^{1/\nu}(\beta-\beta_{\mathrm{c}})] + L^{-\omega} f_1[L^{1/\nu}(\beta-\beta_{\mathrm{c}})]+\cdots$, where $f_0$ and $f_1$ are smooth scaling-functions, the dots stand for sub-dominant corrections to scaling, and $\nu=1$ for this universality class. Defining $\beta_{L,g}$ through $g(L,\beta_{L,g})=f_0(0)$, we obtain the scaling form
\begin{equation}\label{eq:beta_L}
    \beta_{L,g}=\beta_c + {\cal A} L^{-(\frac{1}{\nu}+\omega)}+\cdots\,,
\end{equation}
where $\mathcal{A}$ is a scaling amplitude and the dots indicate sub-leading corrections to scaling. The corrections to scaling in the two-dimensional Ising model universality class have been elucidated using tools from conformal field theory~\cite{caselle:02}. 
\begin{figure}[ht!]
    \includegraphics[width=1.0\linewidth]{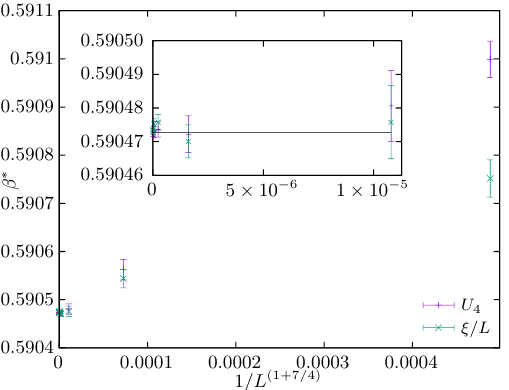}
    \caption{Computation of the critical temperature of the square-lattice Blume-Capel model at $\Delta = 0$. Results based on the Binder cumulant ($U_4$) and the second-moment correlation length over the system size ($\xi_2/L$) are shown. The inset is a zoomed version of the main panel for $L \geq 64$.} 
    \label{fig:Tc_BC}
\end{figure}
For this universality class,  the dominant correction is $\omega = 7/4$, which corresponds to the leading \emph{analytic} corrections to scaling, dominating over the irrelevant renormalization-group eigenvalues the largest of which has $\omega'=2$~\cite{caselle:02}. In fact, both $\beta_{L,U}$ and $\beta_{L,\xi_{2}/L}$, as shown in Fig.~\ref{fig:Tc_BC}, approach their large-$L$ limits \emph{faster} than predicted by the scaling form in Eq.~\eqref{eq:beta_L}. For all our results with $L \geq 64$, we observed compatibility with a constant value (so determining the amplitudes for the $\omega=7/4$ and $\omega'=2$ correction terms is not possible, given our numerical accuracy). As a result, we fit the data to a constant, as the statistical errors for $\beta_{L,U}$ and $\beta_{L,\xi_2/L}$ decrease significantly with increasing $L$. Thus, the fit essentially yields the value for $L = 4096$. This process led to the estimate of $T_{\mathrm{c}}(\Delta = 0)$, as quoted in the main text.

\section{Correlation function analysis} 
\label{sec:cor_function}

Our analysis of the spatial correlations closely follows that of Refs.~\cite{fernandez:18b,fernandez:19,gonzalez-adalid-pemartin:21}, so we provide only a brief review of the key details of our computations. The correlation function $C(r;t)$ was analyzed using the $r$-integrals 
\begin{equation}\label{eq:I_n}
    I_n(t) = \int_0^\infty r^n C(r;t) dr.
\end{equation}
The coherence length was calculated from $\xi(t) = I_2(t)/I_1(t)$. The large system sizes we use enabled the high accuracy of $C(r;t)$, allowing us to reach $\xi$ values up to $1133.3(4)$. Specifically, the integration of Eq.~\eqref{eq:I_n} involved three major steps:
(i) A cut-off distance, $r_{\rm cut}$, is determined such that $C(r_{\rm cut}+1;t)$ is smaller than three times its error, thereby establishing an upper integration cutoff based on the accuracy of our data.
(ii) Next, the maximum of $r^2 C(r;t)$, located at $r^\ast$, is found. From its value, ${r^\ast}^2 C(r^\ast;t)$, the characteristic distances $r_{\rm min}$ and $r_{\rm max}$ are defined as the first $r$ where $r_{\rm min}^2 C(r_{\rm min};t) < 0.9{r^\ast}^2 C(r^\ast;t)$ and $r_{\rm max}^2 C(r_{\rm max};t) < {r^\ast}^2 C(r^\ast;t)/3$, respectively. We require that $r^\ast<r_{\rm min}<r_{\rm max}<r_{\rm cut}$; if this condition is not met, we increase our statistics by performing additional simulations. These characteristic distances are then used to calculate the integrals of Eq.~\eqref{eq:I_n}.
(iii) If $r_{\rm max} - r_{\rm min} \leq 8$, a numerical integration of $C(r;t)$ is performed up to $r_{\rm cut}$. However, if $r_{\rm max} - r_{\rm min} > 8$, $C(r;t)$ is fitted in the interval $[r_{\rm min}, r_{\rm max}]$ to the function
\begin{equation}\label{eq:C_extrapolation}
    F(r) = A \exp{\left[ -(r/\xi_F)^\beta \right]} / \sqrt{r}.
\end{equation}
Next, a numerical integration is carried out up to $r_{\rm max}$, followed by an integration in $F(r)$ from $r_{\rm max}$ to $20\xi_F$. Due to the sparse sampling of $C(r;t)$ with respect to $r$, cubic spline interpolation was applied~\cite{press:92}. Errors were calculated using the jackknife resampling method on the measured correlation functions.

\begin{figure}
    \includegraphics[width=1.0\linewidth]{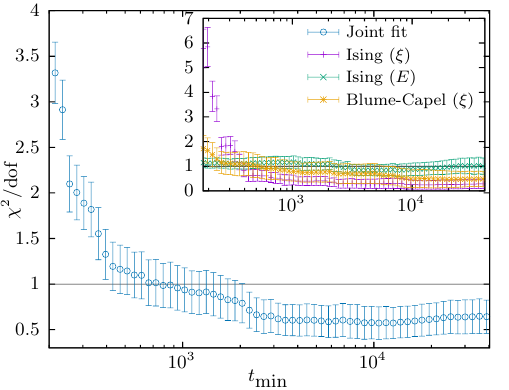}
    \caption{$\chi^2$ over the degrees of freedom for the fits concerning the dynamic critical exponent $z$, as shown in Fig. 3 of the main text. The main panel displays the joint fit of the coherence length and energy for the $L=2^{22}$ Ising model, along with the coherence length for the $L=2^{19}$ Blume-Capel model. The inset shows the respective $\chi^2/{\rm dof}$ values of the individual data sets that were used in the joint fit.}
    \label{fig:chi2}
\end{figure}

\section{Fitting procedures} 
\label{sec:fittings}

For the fitting process, the least squares method was used to calculate the dynamical critical exponent $z$, utilizing the Gnu Scientific Library (GSL)~\cite{GSL}. Since only the diagonal elements of the covariance matrix were considered, a simple fitting approach would fail to account for data correlations and, consequently, the errors. To address this, the independent realizations were analyzed using the jackknife resampling method~\cite{yllanes:11}.

\begin{figure}
    \includegraphics[width=1.0\linewidth]{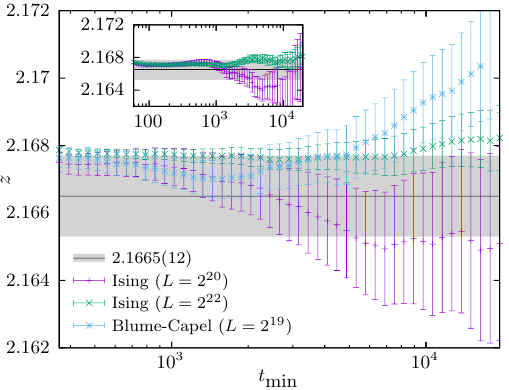}
    \caption{Several estimations of $z$ are shown. The main panel presents the fit results for $\xi$ following Eq.~\eqref{eq:z}, whereas the inset shows $z$ from the excess energy fits based on Eq.~\eqref{eq:z_for_E}. In both panels, the $x$-axis begins at the point where $\chi^2/{\rm dof} \approx 1$ and $Q > 10\%$.}
    \label{fig:z_all}
\end{figure}

Given our data, the exponent $z$ can be calculated using the two methods discussed in the main text, which are reiterated here for the reader's convenience. Starting from the equation $\frac{ d\log{(t)} }{ d\log{(\xi)} } = z + a\xi^{-\omega}$ for a finite system, and integrating, we obtain the following relation for the coherence length 
\begin{equation}\label{eq:z}
    \log{(t)} = z \log{[\xi(t)]} + b + b'\xi(t)^{-\omega}.
\end{equation}
For the energy, we have
\begin{equation}\label{eq:z_for_E}
    E(t) - E_{\text{eq}}= bt^{-(1/z)} + b't^{-(2/z)} + b''t^{-[(\omega+1/\nu)/z]}.
\end{equation}
In Eqs.~\eqref{eq:z} and \eqref{eq:z_for_E}, $b$, $b'$, and $b''$ are fitting parameters (which are obviously different for each equation), $\omega$ is the corrections-to-scaling exponent, and $E_{\text{eq}}$ is the equilibrium value of the energy. Since $E_{\text{eq}}$ factors into the fit, even small deviations can have a significant impact. Therefore, we chose to use this approach only for the Ising model, where $E_{\text{eq}}=-\sqrt{2}$ is exactly known. Specifically, for Eq.~\eqref{eq:z}, since time is the dependent variable, the derivative of the fitting function must also be considered when weighting each measurement in the calculation of $\chi^2$. The weights are then defined as $t_{\rm error} + \xi_{\rm error}f'(\xi)$, where $t_{\rm error}$ and $\xi_{\rm error}$ represent the errors in time and coherence length, respectively, and $f'(\xi)$ is the derivative of the fitting function with respect to $\xi$. Since $t$ is exactly known, we defined it as an infinitesimally small quantity $10^{-40}$. 

Due to the time correlations in our data, we considered fits as acceptable when $\chi^2/{\rm dof} \lesssim 1$, with a fit quality $Q\geq 10\%$. The value of $Q$ can be calculated from $\chi^2$ and ${\rm dof}$ using the regularized upper incomplete gamma function $\gamma(a,x)=\frac{1}{\Gamma(a)}\int_x^\infty t^{a-1} e^{-t} dt$, where $\Gamma$ is the gamma function, since $Q = \gamma({\rm dof}/2,\chi^2/2)$. For the joint fits involving both Eqs.~\eqref{eq:z} and~\eqref{eq:z_for_E}, we defined a piecewise function, that depends on the respective data set.

Figure~\ref{fig:chi2} shows $\chi^2/{\rm dof}$ as a function of the starting fitting range, $t_{\rm min}$, for the fits presented in the main text. The main panel focuses on the joint fit between the $L=2^{22}$ Ising model, using both the coherence length and energy, along with the $\xi$ from the $L=2^{19}$ Blume-Capel model. The inset illustrates $\chi^2/{\rm dof}$ from the individual fits of the data sets that contributed to the joint fit. The errors were computed using the jackknife method. For values of $t_{\rm min}$ above a certain threshold, both $\chi^2/{\rm dof}$ and $Q$ remain consistently in the acceptable range. Figure~\ref{fig:z_all} presents the results for the exponent $z$ across various data sets. The main panel shows fits of the form given by Eq.~\eqref{eq:z}, whereas the inset shows the results from Eq.~\eqref{eq:z_for_E}. The agreement between these results motivated us to pursue a joint fit between the $\xi$ and $E$ data.

\section{Capturing the thermodynamic limit} 
\label{sec:thermo_limit}

\begin{figure}
    \includegraphics[width=1.0\linewidth]{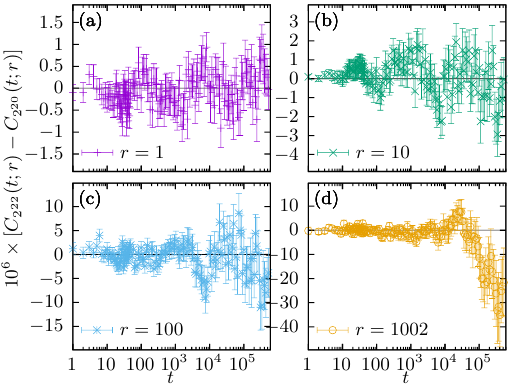}
    \caption{Difference of the correlation function for the Ising model with system sizes $L=2^{20}$ and $L=2^{22}$ over time, for various values of the distance: (a) $r=1$, corresponding to the energy, (b) $r=10$, (c) $r=100$, and (d) $r=1002$.}
    \label{fig:diff}
\end{figure}

In the out-of-equilibrium regime, correlations at a distance $r$ decay exponentially as $[r/\xi(t)]^a$, where the exponent $a > 1$. This suggests that finite-size effects should decay exponentially as
$[L/\xi(t)]^a$. Consequently, significant finite-size effects in out-of-equilibrium dynamics are expected only when the coherence length 
$\xi(t)$ becomes larger than a fraction of the system size $L$. A typical threshold is $1/7$, meaning that finite-size effects are often considered negligible when $\xi(t) < L/7$~\cite{janus:08b}). However, this rule of thumb overlooks a more important consideration: namely, \emph{are the finite-size effects in our simulations significantly smaller than the statistical errors?} In other words, the acceptable size of finite-size effects is not determined \emph{a priori} but is instead set by the statistical errors in the simulation. In our case, these errors are particularly small due to the large system sizes used in our simulations ($L=2^{19}$ for the Blume-Capel model and up to $L=2^{22}$ for Ising model). Given that we limited 
$\xi(t)$ to be no larger than approximately $10^3$ lattice spacings, we are in the range where $L/\xi\approx 4 \times 10^{3}$, which suggests that finite-size effects are unlikely to be significant. Nonetheless, we have performed a quantitative check, which we describe next.

Figure~\ref{fig:diff} shows the difference in the values of 
$C(r,t)$ obtained from simulations for $L=2^{20}$ and $L=2^{22}$, for four values of the spatial separation $r$: the energy ($r = 1$), $r = 10$, $r = 100$, and $r = 1002$. The differences are consistent with zero in all four cases. Notable time correlations are visible in Fig.~\ref{fig:diff}, where the difference remains either positive or negative for a significant period. However, deviations from zero at any given time are well within the statistical errors.

Having found no evidence of finite-size effects in our raw data, we proceeded with a second check. We compared the computation of the 
$z$-exponent from the fits to Eqs.~\eqref{eq:z} and~\eqref{eq:z_for_E} for different system sizes. The results of these fits are shown in Fig.~\ref{fig:z_all}, where the reader can verify that our estimates are indeed not affected by finite-size effects (given the scale of our statistical errors, of course).

\bibliography{biblio}

\end{document}